
\overfullrule=0pt
\def\normalparindent{24pt}

\def\beginparmode{\endmode\begingroup \def\endmode{\par\endgroup}}
\let\endmode=\par
\def\\{\cr}
\def\body{\beginparmode\parindent=\normalparindent}
\def\head#1{\par\goodbreak{\immediate\write16{#1}
    {\noindent\bf #1}\par}\nobreak\nobreak}
\def\refto#1{$[#1]$}
\def\ref#1{Ref.~#1}
\def\refs#1{Refs.~$#1$}
\def\Ref#1{Ref.~$#1$}
\def\cite#1{{#1}}
\def\[#1]{[\cite{#1}]}
\def\Refto#1{$^{\cite{#1}}$}

\def\(#1){(\call{#1})}
\def\call#1{{#1}}
\def\taghead#1{{#1}}
\def\references{\head{REFERENCES}\beginparmode\frenchspacing\parskip=0pt}
\gdef\refis#1{\item{$^{#1}$}}
\gdef\journal#1,#2,#3,#4.{{\sl #1}\ {\bf #2} (#3) #4} 
\gdef\Journal#1,#2,#3,#4.{#1~{\bf #2}, #3 (#4)}
\def\endreferences{\body}
\def\endit{\endmode\vfill\supereject}\let\endpaper=\endit
%
\catcode`@=11
\newcount\r@fcount \r@fcount=0\newcount\r@fcurr
\immediate\newwrite\reffile\newif\ifr@ffile\r@ffilefalse
\def\w@rnwrite#1{\ifr@ffile\immediate\write\reffile{#1}\fi\message{#1}}
\def\writer@f#1>>{}
\def\citeall#1{\xdef#1##1{#1{\noexpand\cite{##1}}}}
\def\cite#1{\each@rg\citer@nge{#1}}
\def\each@rg#1#2{{\let\thecsname=#1\expandafter\first@rg#2,\end,}}
\def\first@rg#1,{\thecsname{#1}\apply@rg}
\def\apply@rg#1,{\ifx\end#1\let\next=\relax%
    \else,\thecsname{#1}\let\next=\apply@rg\fi\next}%
\def\citer@nge#1{\citedor@nge#1-\end-}
\def\citer@ngeat#1\end-{#1}
\def\citedor@nge#1-#2-{\ifx\end#2\r@featspace#1
    \else\citel@@p{#1}{#2}\citer@ngeat\fi}
\def\citel@@p#1#2{\ifnum#1>#2{\errmessage{Reference range #1-#2\space is bad.}
    \errhelp{If you cite a series of references by the notation M-N, then M
    and N must be integers, and N must be greater than or equal to M.}}\else%
    {\count0=#1\count1=#2\advance\count1 by1\relax\expandafter\r@fcite\the%
    \count0,\loop\advance\count0 by1\relax 
    \ifnum\count0<\count1,\expandafter\r@fcite\the\count0,%
    \repeat}\fi}
\def\r@featspace#1#2 {\r@fcite#1#2,}
\def\r@fcite#1,{\ifuncit@d{#1}\expandafter\gdef\csname r@ftext\number%
    \r@fcount\endcsname {\message{Reference #1 to be supplied.}%
    \writer@f#1>>#1 to be supplied.\par }\fi\csname r@fnum#1\endcsname}
\def\ifuncit@d#1{\expandafter\ifx\csname r@fnum#1\endcsname\relax%
    \global\advance\r@fcount by1%
    \expandafter\xdef\csname r@fnum#1\endcsname{\number\r@fcount}}
\let\r@fis=\refis
\def\refis#1#2#3\par{\ifuncit@d{#1}%
    \w@rnwrite{Reference #1=\number\r@fcount\space is not cited up to now.}%
    \fi\expandafter\gdef\csname r@ftext\csname r@fnum#1\endcsname\endcsname%
    {\writer@f#1>>#2#3\par}}
\def\r@ferr{\endreferences\errmessage{I was expecting to see
    \noexpand\endreferences before now; I have inserted it here.}}
\let\r@ferences=\references
\def\references{\r@ferences\def\endmode{\r@ferr\par\endgroup}}
\let\endr@ferences=\endreferences
\def\endreferences{\r@fcurr=0{\loop\ifnum\r@fcurr<\r@fcount
    \advance\r@fcurr by
1\relax\expandafter\r@fis\expandafter{\number\r@fcurr}%
    \csname r@ftext\number\r@fcurr\endcsname%
    \repeat}\gdef\r@ferr{}\endr@ferences}
\let\r@fend=\endpaper
\gdef\endpaper{\ifr@ffile\immediate\write16{Cross References
    written on []\jobname.REF.}\fi\r@fend}
\catcode`@=12
\citeall\refto\citeall\ref\citeall\Ref\citeall\refs

\catcode`@=11
\newcount\tagnumber\tagnumber=0
\immediate\newwrite\eqnfile\newif\if@qnfile\@qnfilefalse
\def\write@qn#1{}\def\writenew@qn#1{}
\def\w@rnwrite#1{\write@qn{#1}\message{#1}}
\def\@rrwrite#1{\write@qn{#1}\errmessage{#1}}
\def\taghead#1{\gdef\t@ghead{#1}\global\tagnumber=0}
\def\t@ghead{}
\expandafter\def\csname @qnnum-3\endcsname
    {{\t@ghead\advance\tagnumber by -3\relax\number\tagnumber}}
\expandafter\def\csname @qnnum-2\endcsname
    {{\t@ghead\advance\tagnumber by -2\relax\number\tagnumber}}
\expandafter\def\csname @qnnum-1\endcsname
    {{\t@ghead\advance\tagnumber by -1\relax\number\tagnumber}}
\expandafter\def\csname @qnnum0\endcsname
    {\t@ghead\number\tagnumber}
\expandafter\def\csname @qnnum+1\endcsname
    {{\t@ghead\advance\tagnumber by 1\relax\number\tagnumber}}
\expandafter\def\csname @qnnum+2\endcsname
    {{\t@ghead\advance\tagnumber by 2\relax\number\tagnumber}}
\expandafter\def\csname @qnnum+3\endcsname
    {{\t@ghead\advance\tagnumber by 3\relax\number\tagnumber}}
\def\equationfile{\@qnfiletrue\immediate\openout\eqnfile=\jobname.eqn%
    \def\write@qn##1{\if@qnfile\immediate\write\eqnfile{##1}\fi}
    \def\writenew@qn##1{\if@qnfile\immediate\write\eqnfile
    {\noexpand\tag{##1} = (\t@ghead\number\tagnumber)}\fi}}
\def\callall#1{\xdef#1##1{#1{\noexpand\call{##1}}}}
\def\call#1{\each@rg\callr@nge{#1}}
\def\each@rg#1#2{{\let\thecsname=#1\expandafter\first@rg#2,\end,}}
\def\first@rg#1,{\thecsname{#1}\apply@rg}
\def\apply@rg#1,{\ifx\end#1\let\next=\relax%
    \else,\thecsname{#1}\let\next=\apply@rg\fi\next}
\def\callr@nge#1{\calldor@nge#1-\end-}
\def\callr@ngeat#1\end-{#1}
\def\calldor@nge#1-#2-{\ifx\end#2\@qneatspace#1 %
    \else\calll@@p{#1}{#2}\callr@ngeat\fi}
\def\calll@@p#1#2{\ifnum#1>#2{\@rrwrite{Equation range #1-#2\space is bad.}
    \errhelp{If you call a series of equations by the notation M-N, then M
    and N must be integers, and N must be greater than or equal to M.}}\else%
    {\count0=#1\count1=#2\advance\count1 by1\relax\expandafter\@qncall\the%
    \count0, \loop\advance\count0 by1\relax%
    \ifnum\count0<\count1,\expandafter\@qncall\the\count0, \repeat}\fi}
\def\@qneatspace#1#2 {\@qncall#1#2,}
\def\@qncall#1,{\ifunc@lled{#1}{\def\next{#1}\ifx\next\empty\else
    \w@rnwrite{Equation number \noexpand\(>>#1<<) has not been defined yet.}
    >>#1<<\fi}\else\csname @qnnum#1\endcsname\fi}
\let\eqnono=\eqno
\def\eqno(#1){\tag#1}
\def\tag#1$${\eqnono(\displayt@g#1 )$$}
\def\aligntag#1\endaligntag $${\gdef\tag##1\\{&(##1 )\cr}\eqalignno{#1\\}$$
    \gdef\tag##1$${\eqnono(\displayt@g##1 )$$}}

\def\eqalignno#1{\displ@y \tabskip\centering
    \halign to\displaywidth{\hfil$\displaystyle{##}$\tabskip\z@skip
    &$\displaystyle{{}##}$\hfil\tabskip\centering
    &\llap{$\displayt@gpar##$}\tabskip\z@skip\crcr #1\crcr}}
\def\displayt@gpar(#1){(\displayt@g#1 )}
\def\displayt@g#1 {\rm\ifunc@lled{#1}\global\advance\tagnumber by1
    {\def\next{#1}\ifx\next\empty\else\expandafter
    \xdef\csname @qnnum#1\endcsname{\t@ghead\number\tagnumber}\fi}%
    \writenew@qn{#1}\t@ghead\number\tagnumber\else
    {\edef\next{\t@ghead\number\tagnumber}%
    \expandafter\ifx\csname @qnnum#1\endcsname\next\else%
    \w@rnwrite{Equation \noexpand\tag{#1} is a duplicate number.}\fi}%
    \csname @qnnum#1\endcsname\fi}
\def\ifunc@lled#1{\expandafter\ifx\csname @qnnum#1\endcsname\relax}
\let\@qnend=\end
\gdef\end{\if@qnfile\immediate\write16{Equation numbers
    written on []\jobname.EQN.}\fi\@qnend}
\catcode`@=12

\font\bigbf=cmbx12 
\magnification=\magstep1
\vsize=23.5truecm\voffset=-0.5truecm 
\baselineskip=16pt
\raggedbottom
\fontdimen16\tensy=2.7pt\fontdimen17\tensy=2.7pt
\abovedisplayskip=6pt plus 3pt minus 3pt
\belowdisplayskip=6pt plus 3pt minus 3pt
\def\section#1#2{\goodbreak \vskip0.6cm \leftskip=20pt \parindent=-20pt
  \indent{\bf\hbox to 20pt{#1\hss}#2} \medskip\nobreak \leftskip=0pt
  \parindent=20pt} 
\def\dd{{\rm d}}\def\ee{{\rm e}}\def\ii{{\rm i}}\def\de{\partial}
\def\eps{\varepsilon}\def\ts#1{{\textstyle{#1}}}
\def\half{\ts{1\over2}}\def\fourth{\ts{1\over4}}
\def\sixth{\ts{1\over6}}\def\eighth{\ts{1\over8}}

\null
\vfill
\centerline{\bigbf Chaos in Robertson-Walker Cosmology}
\bigskip
\centerline{Luca Bombelli\footnote{*}{E-mail: luca@beauty1.phy.olemiss.edu}}
\smallskip
\centerline{\sl Department of Physics and Astronomy,}
\centerline{\sl University of Mississippi, University, MS 38677, U.S.A.}
\medskip
\centerline{Fernando Lombardo\footnote{\dag}
{E-mail: lombardo@df.uba.ar}}
\smallskip
\centerline{\sl Departamento de F\'{\i}sica, Facultad de Ciencias Exactas
y Naturales,}
\centerline{\sl Universidad de Buenos Aires, Ciudad Universitaria,
Buenos Aires, Argentina}
\medskip
\centerline{Mario Castagnino\footnote{\ddag}{E-mail: castagni@iafe.uba.ar}}
\smallskip
\centerline{\sl Instituto de Astronom\'{\i}a y F\'{\i}sica del Espacio,}
\centerline{\sl Casilla de Correos 67, Sucursal 28, 1428 Buenos Aires,
Argentina.}
\vfill
\centerline{Abstract}
\midinsert\narrower
\noindent Chaos in Robertson-Walker cosmological models where gravity is
coupled to one or more scalar fields has been studied by a few authors, mostly
using numerical simulations. In this paper we begin a systematic study of the
analytical aspect. We consider one conformally coupled scalar field and, using
the fact that the model is integrable when the field is massless, we show in
detail how homoclinic chaos arises for nonzero masses using a perturbative
method.
\endinsert
\vfill
PACS numbers: 98.80.Hw, 04.40.Nr


\vfill\eject

\section{I.}{Introduction}\taghead{1.}

\noindent In \ref{Cal1}, Calzetta and El Hasi presented an argument,
supported by numerical evidence, for the appearance of chaotic behaviour in
a spatially closed Robertson-Walker (RW) cosmological model, filled with a
conformally coupled massive scalar field. Because cosmological models of this
type are practically the simplest possible ones, but give in some sense a
rough, smoothed out indication of the average properties of a more realistic
cosmology, they have frequently been used as a first testing ground on which
to try out approaches to both classical and quantum problems in cosmology.
In addition, because of their kinematical simplicity, one would expect the
dynamics of these models to be equally simple. It thus may come somewhat as a
surprise that chaos should appear at this level, although it is known that it
can appear in a nonlinear dynamical system with two degrees of fredom, and one
may suspect that it is in fact generic for relativistic systems.

The study of chaos in relativistic cosmology began more than twenty years
ago with the study by the Russian school\Refto{BKL} and Misner's
group\Refto{Misn} on diagonal Bianchi type IX (mixmaster) vacuum cosmologies.
This research is, by itself, very interesting and still active,\Refto{Pull}
but the hopes initially placed in these models, essentially the solution of
the horizon problem from the mixing effect associated with chaos, could not be
sustained, and the Bianchi IX cosmologies have been mostly used as toy models,
rather than first approximations from which to extract estimates of physical
quantities. On the other hand, although the question of which RW model best
approximates our inhomogeneous, anisotropic universe is not yet settled, there
is agreement on the relevance of RW models both for classical and quantum
cosmology,\Refto{Berr} and physical predictions are often made based on them.
Therefore the presence of chaos, in this kind of cosmologies, enlarges the
avenues of research in fundamental subjects such as the arrow of time and the
issue of irreversibility in the evolution of the universe, or the transition
from the quantum regime to the classical one, with the corresponding
appearance of decoherence and correlations and related phenomena like particle
creation, and has been used in considerations related to physical phenomena in
inflationary universes.\Refto{Cal3,Corn,Fine}

In their work, Calzetta and El Hasi used mostly a numerical method to
investigate the chaotic behaviour of the model, but also sketched an
analytical, perturbative method by which they considered the model as a
perturbation of the integrable one obtained with a vanishing mass for the
scalar field, and split the Hamiltonian of the system into an integrable
part and a coupling term. With a perturbative argument commonly used in the
treatment of near-integrable systems, based on the Melnikov criterion for
homoclinic chaos and Chirikov's resonance overlap criterion,\Refto{Zasl}
they argued qualitatively that the KAM tori of the integrable part are
destroyed and replaced by stochastic layers under the effect of the
perturbation. The resulting strong indications of chaotic behaviour were
confirmed by the numerical analysis, which allowed them to go beyond
perturbation theory.

The generic transition to chaos in a perturbed integrable Hamiltonian system
(``soft chaos") can be seen as taking place in two steps.\Refto{Zasl,Gutz} In
the first one, a perturbation of the Hamiltonian, with a nonvanishing Fourier
component in resonance with one of the tori of the unperturbed system,
modifies the dynamics in a neighborhood of this torus to give a new integrable
system, which however has heteroclinic orbits even if before there were none.
These orbits are separatrices, which act as seeds of chaos when in the second
step one takes into account the effect of the remaining components of the
perturbation on the new integrable dynamics; it is here that one may use the
Melnikov method to determine whether in effect the dynamics becomes chaotic.
What this method provides is a topological criterion for detecting a chaotic
``tangle'' in the orbits near the separatrices, and is therefore free of the
coordinate ambiguities that have arisen, e.g., in the study of the chaotic
nature of Bianchi models by other methods; for a quick introduction to the
Melnikov method see, e.g., \refs{Ozor} or \cite{Bom1}, and \ref{Guck} for a
more technical exposition.

The goal of this paper is to continue the work in \ref{Cal1} by starting to
examine in detail the analytical treatment of the model; in future work, we
will systematically extend our work to more general models. We will use a
different perturbative expansion of the Hamiltonian, and show the onset of
chaotic behavior by an explicit computation of the Melnikov integral. Other
recent work on chaos in Robertson-Walker models coupled to scalar fields has
focused on their application to inflation,\Refto{Corn} again using mostly
numerical methods, and an analysis based on the Painlev\'e analysis of
differential equations,\Refto{Vuce} which confirmed the non-integability of
the models. The main advantage of our work is the amount of information it can
give us regarding the resonances responsible for the onset of chaos and the
chaotic region of phase space.

The paper is organized as follows: in section II, we set up the perturbative
method we will use, by choosing appropriate variables to describe the
cosmological model and separating the corresponding Hamiltonian into an
``unperturbed" term and a ``perturbation;" in section III we analyze the first
step above for our example, i.e., we discuss the local dynamics in the
vicinity of one of the resonant tori, while in section IV we develop the second
step, discussing the effect of other perturbation terms on the local
dynamics, which becomes chaotic as Melnikov's method shows; section V contains
concluding remarks on possible future developments.


\section{II.}{Dynamical variables and Hamiltonian:\hfil\break
setup for the perturbative treatment}\taghead{2.}

\noindent Let us consider the closed Robertson-Walker (RW) metric for a
homogeneous and isotropic universe, written in the form
$$
   \dd s^2 = a^2(t) [ -\dd t^2 + \dd\chi^2 + \sin^2\!\chi
   (\dd\theta^2 + \sin^2\!\theta\,\dd\varphi^2) ] \;, \eqno(Metric)
$$
where $0\le\varphi\le2\pi$, $0\le\theta\le\pi$, $0\le\chi\le\pi$ are the
angular coordinates on S$^3$, and $t$ is the usual ``conformal time" (see,
e.g., \ref{MTW}).  We shall consider only models possessing a cosmic
singularity $a = 0$, chosen for convenience to occur at $t = 0$. We shall also
assume that after the big crunch a new cosmological cycle begins, but $a$
changes sign, in such a way that the evolution is smooth, and this process is
repeated an infinite number of times. This well-known possibility of extending
the evolution beyond the $a=0$ singularity is an artifact of the symmetry
possessed by the RW models (we have no grounds for believing that it can be
generalized), but it is convenient for us because it allows us to assume that
time runs up to $t=+\infty$, and thus to use techniques for detecting chaos,
including its very definition, that would otherwise be inapplicable. Any
conclusion we draw from the model would of course not have to rely on this
extension; the time scale for any chaotic behavior to set in must be shorter
than one cycle for its physical consequences to be meaningful.

We start with the Einstein-Hilbert action for the gravitational field,
$$
   S_{\rm g}[g] = {1\over16\pi G}\int \dd^4x\,\sqrt{-g}\,R \;,
   \eqno(ActionGr)
$$
and the action
$$
   S_{\rm f}[\Phi,g] = -\half \int \dd^4x\, \sqrt{-g}\,
   \left[g^{ab}\nabla\!_a\Phi\,\nabla\!_b\Phi
   + (\mu^2+\sixth\,R)\,\Phi^2\right] \eqno(ActionSc)
$$
for a real, conformally coupled scalar field of mass $\mu$. To write
down the Hamiltonian, use the fact that the square root of the
determinant of the metric \(Metric) and its scalar curvature are,
respectively,
$$
   \sqrt{-g} = a^4\sin^2\!\chi\,\sin\theta \;, \qquad
   R = 6 \left({\ddot a\over a^3}+{1\over a^2}\right), \eqno(Curvature)
$$
where an overdot denotes a $t$-derivative, and reparam\-etrize the scalar
field (which must be homogeneous for consistency) by $\Phi \mapsto \phi:=
\sqrt{4\pi G/3}\,a\Phi$. Then we obtain, up to a constant overall factor,
$$
   H(a,\phi;\pi,p)
   = \half\,\big[-(\pi^2+a^2)+(p^2+\phi^2)+\mu^2a^2\phi^2\big]\;,
   \eqno(InitialH)
$$
where $\pi$ and $p$ are the momenta conjugate to $a$ and $\phi$,
respectively. Here, as usual with general relativistic systems, time
reparametrization invariance requires that we impose that all solutions
satisfy the constraint
$$
   H = 0 \;, \eqno(Constraint)
$$
which is just the $(0,0)$-component of the Einstein equation.

The system described by the Hamiltonian \(InitialH) consists of two harmonic
oscillators (of which one is ``inverted," as is to be expected from any
degree of freedom related to the spatial volume element), coupled through
a term proportional to $\mu^2$. For $\mu^2=0$, the system is trivially
integrable; our main goal is to analyze the effect of the coupling term,
at least for small values of $\mu^2$. However, in view of the perturbative
treatment to be carried out, the form \(InitialH) has two disadvantages: the
``perturbation" term $\mu^2a^2\phi^2$ is not always small, in the sense that
for a fixed value of $\mu^2$ it cannot be uniformly bounded for all orbits
of the unperturbed system; and the latter system is degenerate, in the sense
that the frequencies of both oscillators are constant. We will not attempt
to deal directly with the first inconvenience, which at any rate is not
really a problem for perturbations of a fixed orbit; we will instead perform
a change of variables that will allow us to isolate a different integrable
Hamiltonian from $H$, with enough of the degeneracy removed for us to be
able to use the Melnikov method of detecting homoclinic chaos. Regarding
the perturbation, it will be sufficient for us to know that it is a
well-behaved function on each trajectory.

Following \ref{Cal1}, we first replace the dynamical variables $p$ and
$\phi$ by new variables $j$ and $\varphi$, respectively, defined as
$$
   \phi = \sqrt{{2j\over\omega}}\,\sin\varphi\;, \qquad
   p = \sqrt{2\,\omega j}\,\cos\varphi \;, \eqno(IntermediateVar1)
$$
where $\omega = \sqrt{1+\mu^2a^2}$ is the instantaneous frequency of the
field; thus, this is a first approximation to action-angle variables for
the scalar field, which takes into account the coupling term, but not the
fact that $a$ is a dynamical variable, and it is a key trick in this
treatment of the model. To complete the canonical transformation, we must
introduce a new momentum variable conjugate to $a$, given by
$$
   P := \pi - {\mu^2aj\over 2\,\omega^2}\sin2\varphi \;,
   \eqno(IntermediateVar2)
$$
and arising from the fact that $\omega$ is $a$-dependent. Now the
Hamiltonian in terms of the new variables can be written as a sum
$H = \hat H_0 + \delta\hat H$, of an unperturbed Hamiltonian
$$
   \hat H_0(a,\varphi;P,j) = -\half\, (P^2 + a^2) + j \sqrt{1+\mu^2a^2} \;,
   \eqno(IntermediateH)
$$
which is obviously integrable, since it has the two commuting constants
of the motion $\hat H_0$ and $j$, and a perturbation
$$
   \delta\hat H(a,\varphi;P,j) = -{\mu^2aPj\over 2\,(1+\mu^2a^2)} \sin2\varphi
   - \left[{\mu^2aj\over 4\,(1 + \mu^2a^2)}\right]^2 (1-\cos4\varphi) \;.
   \eqno(IntermediateDeltaH)
$$
However, the system \(IntermediateH) is not easy to integrate; since we
will need to perform explicit calculations, and in particular to write it
in terms of action-angle variables, we find it convenient to simplify it
somewhat.

In \ref{Cal1}, this simplification was achieved by noticing that for most of
the unperturbed motion, the value of $\omega$ is very large, and using the
$\omega\gg1$ approximation to $\hat H_0$. In addition, because the treatment
was a perturbative one in the parameter $\mu^2$, or more precisely $\mu^2j$,
the second term in the Hamiltonian perturbation \(IntermediateDeltaH) was
dropped, since it contains higher powers of $\mu^2$ than the first one.
Notice that, implicit in this way of proceeding is the idea that one
considers the $\mu^2$ in $\hat H_0$ as a {\it fixed\/} parameter, while in
$\delta\hat H$, $\mu^2=:\eps$ is rendered temporarily independent of the
first one for the sake of the perturbative treatment, and allowed to vary in
a neighborhood of $\mu^2=\eps=0$. (One may be able to set up a perturbation
theory for Hamiltonians of the type $H_0(q,p;\eps) + \eps\,\delta H(q,p)$,
where $H_0(q,p;\eps)$ is integrable for all $\eps$, but to our knowledge
such a theory is not available yet.)

Here, we will follow a different path. Since we are going to expand the
perturbation in powers of $\mu^2$ and neglect terms higher than $\mu^4$, we
will do the same for $\hat H_0$, and retain there only the terms up to order
$\mu^2$ while we will put in the perturbation the higher order ones. The new
unperturbed system obtained in this way will be explicitly integrable. As the
perturbation parameter, we will use $\mu^2$, rather than the dimensionless
$\mu^2j$; this is mainly to avoid having to explain already at this stage
what values of the variable $j$ we are interested in. At any rate, strictly
speaking we will not have to talk about $\eps=\mu^2$ being ``small" (if we
did, we might say that $\mu/\mu_{\rm P}\ll1$, where $\mu_{\rm P}$ is the
Planck mass); the perturbative results we will obtain are valid for $\eps$
in some interval near zero, and such topological statements are
``dimensionless." Expanding $\sqrt{1 + \mu^2a^2} = 1 + \half\,\mu^2a^2 -
\fourth\,\mu^4a^4 + o(\mu^4)$, we obtain a new breakup of the Hamiltonian
into
$$
   H = H_0 + \delta H\;, \eqno(Hamiltonian)
$$
where the unperturbed part is now
$$
   H_0(a,\varphi;P,j) = -\half\,(P^2+a^2) + j\,(1+\half\,\mu^2a^2) \;,
   \eqno(NewIntermediateH)
$$
and the perturbation differs from \(IntermediateDeltaH) by terms proportional
to $\mu^4 a^4$ and of higher order in $\mu^2$, namely
$$
   \delta H = -\half\,\mu^2aPj\,\sin 2\varphi
   + \mu^4 \left[ \half\,a^3 Pj\,\sin 2\varphi - \fourth\,a^4j
   - \ts{1\over16}\,a^2j^2 \,(1-\cos 4\varphi) \right] + o(\mu^4) \;.
   \eqno(NewPert)
$$

One of the action variables of the new unperturbed system is $j$ itself; to
find the other one, we compute the integral $k:=(2\pi)^{-1}\int P\,\dd a$,
along the trajectory specified by the values of the conserved quantities $j$
and $H_0 = {\rm const} =: h_0$ (ultimately, we will be interested only in
trajectories near $h_0=0$, in order to satisfy the constraint $H_0+\delta H
= 0$, but for now we keep the treatment more general), for which
$$
   P(a) = \pm\, \sqrt{2\,(j-h_0) - (1-\mu^2 j)\,a^2} \;. \eqno(UnperturbedP)
$$
The integration yields
$$
   k = {j-h_0 \over \sqrt{1 - \mu^2j}} \;, \eqno(k)
$$
and the angle variables canonically conjugate to $k$ and $j$, respectively,
are given by
$$ \eqalignno{
   \theta &= \arctan \left(\sqrt{1-\mu^2j}\,a/P\right) &(Theta) \cr
   \delta &= \varphi - {\mu^2aP \over 4\,(1-\mu^2j)} \;, &(Delta)}
$$
as can be checked by direct computation of the Poisson brackets. Notice that
$\mu^2$ here is not the infinitesimal $\eps$, but we do want to consider it
as being small, so for $j$ not too big $1-\mu^2j > 0$. With these new
variables we can now write down our final expression for the unperturbed
Hamiltonian,
$$
   H_0(k,j) = j - k\,\sqrt{1-\mu^2j} \;. \eqno(finalH)
$$
Choosing a value for $h_0$ will then impose a relationship between $j$ and
$k$; in particular, for $h_0\approx0$, since $j$ is positive---see
\(NewIntermediateH)---$k$ must be positive as well. For the perturbation, we
invert the transformation \(k)--\(Delta), and substitute in \(NewPert), which,
keeping only the lowest order terms in $\eps=\mu^2$, gives
$$ \eqalignno{
   \delta H
   =\ &\fourth\,\eps\,kj \Bigl[\cos(2\theta+2\delta)
   - \cos(2\theta-2\delta) \Bigr]
   + \fourth\,\eps^2 \Bigl[ -\ts{3\over2}\,k^2j - \fourth\,kj^2\ + \cr
   &+ (2\,k^2j + \fourth\,kj^2) \cos2\theta - \half\,k^2j \cos4\theta
   - \half\,k^2j \cos(2\delta-\pi/2) + \fourth\,kj^2 \cos4\delta\ + \cr
   \noalign{\smallskip}
   &+ k^2j \cos(2\theta-2\delta) - k^2j \cos(2\theta+2\delta)
   - \eighth\,kj^2 \cos(2\theta+4\delta)
   - \eighth\,kj^2 \cos(2\theta-4\delta)\ + \cr
   &+ \ts{\sqrt5\over4}\,k^2j \cos(4\theta+2\delta+\psi)
   - \ts{\sqrt5\over4}\,k^2j \cos(4\theta-2\delta+\psi)\Bigr]
   + O(\eps^3)\;, &(finalDeltaH)}
$$
where $\psi:=-\arcsin(1/\sqrt5)$, and we have chosen to write the various
terms in the form that will be most useful in the following. This completes
the setup for the perturbative treatment of the model. The dynamics of the
unperturbed Hamiltonian $H_0$ is trivial, and gives a conditionally periodic
motion $\theta=\theta_0+\omega_k t$, $\delta=\delta_0+\omega_j t$, with
frequencies given by
$$ \eqalign{
   \omega_k &= {\partial H_0\over\partial k} = -\sqrt{1-\mu^2j} \cr
   \omega_j &= {\partial H_0\over\partial j}
   = {\mu^2k + 2\,\sqrt{1-\mu^2j} \over 2\,\sqrt{1-\mu^2j}} \;. }
   \eqno(Frequencies)
$$
In the next section we will start to take a look at the perturbed dynamics.


\section{III.}{Effect of the resonant terms in the perturbation}\taghead{3.}

\noindent The effect of the perturbation $\delta H$ is felt in particular
on the rational tori of the unperturbed dynamics, the ones where the motion
becomes periodic, because the frequencies \(Frequencies) associated with the
two degrees of freedom are related by the resonance condition
$$
   n_0\, \omega_k + m_0\, \omega_j = 0\;, \eqno(Resonance1)
$$
for some integer numbers $n_0$ and $m_0$. Using the explicit form of the
frequencies in the resonance condition, we obtain a relationship between
the action variables $k$ and $j$, namely that
$$
   {2\,(1 - \mu^2 j) \over \mu^2 k + 2\,\sqrt{1 - \mu^2j}} = {m_0\over n_0}
   \eqno(Resonance2)
$$
must be a rational number. The latter equation can be expressed as
$$
   \mu^2 k = 2\,{n_0\over m_0}\,(1-\mu^2j) - 2\,\sqrt{1-\mu^2j}\;, \eqno(jk)
$$
and, since $k$ must be positive in order for $h_0\approx0$, we see that
a necessary condition is that $n_0$ and $m_0$ have the same sign---we can
take them to be positive---, with $n_0$ strictly greater than $m_0$.

To find out which of the resonant tori of the unperturbed system are affected
by $\delta H$, we must consider the perturbation \(finalDeltaH) as a Fourier
series with respect to the angular variables $\theta$ and $\delta$, of the
form
$$ \eqalignno{
   &\delta H(\theta,\delta;k,j)
   = \sum_{i=1}^\infty\,\delta H^{(i)}(\theta,\delta;k,j) \cr
   &\delta H^{(i)}(\theta,\delta;k,j)
   = \eps^i \sum_{n,m}\,V_{nm}^{(i)}(k,j)
   \,\cos(n\theta+m\delta+\psi_{nm}^{(i)})\;, &(FourierSeries)}
$$
and check which of the Fourier coefficients $V^{(1)}_{nm}(k,j)$ are
non-vanishing. The order $\eps$ terms in \(finalDeltaH) are
$$
   \delta H^{(1)} = \fourth\,\eps\,kj
   \Bigl[\cos(2\theta+2\delta) - \cos(2\theta-2\delta) \Bigr].
   \eqno(DeltaHone)
$$
Therefore, to order $\mu^2$ in the perturbation, the resonant tori which are
broken by the perturbation are those for which
$$
   n_0 = \pm\, m_0 = 2 \;.
$$
However, neither of these pairs satisfies the conditions given below \(jk),
which means that these resonant tori are not present in the physically
relevant region near $h_0 = 0$.

This forces us to consider the Fourier components of order $\eps^2$ in the
perturbation \(finalDeltaH). In order to look at their effect on the
dynamics, however, we must somehow consider them as perturbations of a system
which includes the order $\eps$ terms. The dynamical behavior of such a
system is qualitatively similar to that of the unperturbed one in the region
near $h_0 = 0$, because $H_0$ has no tori there which resonate with the order
$\eps$ terms, and to set up the system we can use the normal techniques
for non-resonant perturbations of integrable systems.

The method (see, e.g., \ref{Gall}, \S5.10) consists in performing an
$\eps$-dependent canonical transformation to new coordinates $(\theta',
\delta'; k',j')$, in which the new Hamiltonian has no perturbation terms of
order $\eps$, and is obtained by using $k'\theta + j'\delta +
\Phi(\theta,\delta; k',j')$ as generating function, where
$$
   \Phi(\theta,\delta; k',j') = -\eps \sum_{n,m} {V_{nm}^{(1)}(k',j')\over
   n\,\omega_k(k',j')+m\,\omega_j(k',j')}\,\sin(n\theta+m\delta+\psi_{nm})\;.
   \eqno(GenFun)
$$
While the new canonical coordinates cannot be explicitly expressed in terms
of the old ones, they differ by a series of powers of $\eps$, and for our
purposes it will be sufficient to calculate the term of order $\eps$. A
calculation shows that the new unperturbed Hamiltonian $H'_0(k',j') =
H_0(k',j')$, i.e., it is given by the same function \(finalH) as in the
previous variables, while the perturbation is now of the form
$$ \eqalignno{
   \delta H' = \eps^2 &\Bigl[
   V_{00}^{\prime(2)}(k',j') + V_{20}^{\prime(2)}(k',j')\,\cos 2\theta'
   + V_{02}^{\prime(2)}(k',j')\,\cos(2\delta'-\pi/2)\ + \cr
   &+ V_{40}^{\prime(2)}(k',j')\,\cos 4\theta'
   + V_{04}^{\prime(2)}(k',j')\,\cos 4\delta'\;+ \cr
   \noalign{\smallskip}
   &+ V_{22}^{\prime(2)}(k',j')\,\cos(2\theta'+2\delta')
   + V_{2-2}^{\prime(2)}(k',j')\,\cos(2\theta'-2\delta')\ + \cr
   \noalign{\smallskip}
   &+ V_{24}^{\prime(2)}(k',j')\,\cos(2\theta'+4\delta')
   + V_{2-4}^{\prime(2)}(k',j')\,\cos(2\theta'-4\delta')\ + \cr
   \noalign{\smallskip}
   &+ V_{42}^{\prime(2)}(k',j')\,\cos(4\theta'+2\delta'+\psi)
   + V_{4-2}^{\prime(2)}(k',j')\,\cos(4\theta'-2\delta'+\psi) \ + \cr
   &+ V_{44}^{\prime(2)}(k',j')\,\cos(4\theta'+4\delta')
   + V_{4-4}^{\prime(2)}(k',j')\,\cos(4\theta'-4\delta') \Bigr]
   + O(\eps^3)\;. &(newDeltaH)}
$$
Since the unperturbed Hamiltonian is the same as before, the condition for a
resonance at $h_0\approx0$ is still $n_0>m_0>0$, and the only term in
\(newDeltaH) satisfying this condition has
$$
   n_0 = 4\;,\quad m_0 = 2\;,
$$
in which case
$$
   \mu^2 k_0 = 2\,\sqrt{1-\mu^2 j_0}\,\bigl(2\,\sqrt{1-\mu^2 j_0}-1\bigr),
   \eqno(resmu4)
$$
is positive if $\mu^2j_0<3/4$. We thus choose to study the motion near the
resonant torus with action variables $(j_0,k_0)$ characterized by the integer
numbers $(n_0,m_0) = (4,2)$, and for which the perturbation coefficient is
$$
   V_{42}^{\prime(2)}(k',j') = V_{42}^{(2)}(k',j')
   = \ts{\sqrt5\over16}\,k^{\prime2}j'. \eqno(Vcoeff)
$$
(It is not necessary for us to write down explicitly all the functions
$V_{mn}^{\prime(2)}(k',j')$ here, but some of them coincide with the
$V_{mn}^{(2)}(k',j')$ in \(finalDeltaH).) As far as this local dynamics is
concerned, then, we could have used directly the Hamiltonian $H_0(k,j)$ with
perturbation $\delta H^{(2)}(\theta,\delta;k,j)$ as in \(finalDeltaH), without
worrying about the presence of the $\eps$ term. From now on, we will drop the
primes on the variables just introduced.

Since the perturbation with coefficient \(Vcoeff) is resonant, it does change
the local dynamics qualitatively near the chosen torus. Following the standard
procedure for such a case (see, e.g. \ref{Zasl}), we go over to a set of
canonical variables $(\gamma,\delta;K,J)$ adapted to the resonant torus,
chosen so that one of the momenta will still be a constant of the motion under
the resonant term in the perturbation:
$$ \eqalign{
   K &:= {k - k_0\over n_0} = \fourth\,(k-k_0)
   \hskip40pt \gamma:= n_0\,\theta + m_0\,\delta +\psi_0
   = 4\,\theta + 2\,\delta + \psi_0 \cr
   J &:= -{m_0\over n_0}\,k+j = -\half\,k + j \cr} \eqno(LocalVariables)
$$
where $\psi_0$ is some arbitrary, fixed angle.

If we suppose that $K\ll1$ is a small increment of the variable $k$ around the
resonant value $k_0$, then we can expand the Hamiltonian, written in the new
variables, in powers of $K$, and study the dynamics generated by the leading
terms. We begin with the resonant part of the perturbation,
$$ \eqalignno{
   \delta H^{(2)} &= \eps^2\,V_{42}^{(2)}(k,j) \cos(4\theta+2\delta+\psi)
   + \hbox{(terms with different $n$ and $m$)} \cr
   &= \ts{\sqrt5\over16}\,\eps^2k_0^2\,j\,\cos(\gamma+\psi-\psi_0)
   + \hbox{(terms with different $n$ and $m$)}\;. &(localDeltaH)}
$$
Here, $j$ is to be thought of as $j(K\!=\!0,J)$, with $J$ arbitrary.
Since this term in the perturbation depends only on $\gamma$ and not on
$\delta$, $J$ is still a constant of the motion; for simplicity we fix its
value at the resonant one, $J_0$. Then for the unperturbed Hamiltonian we
obtain
$$
   H_0(K,J_0) = H_0(k_0,j_0) + \half\,\Omega\,K^2
   + O(K^3) \;, \eqno(localH_0)
$$
where
$$
   \Omega:= {\partial^2 H_0\over\partial k^2}\,n_0^2
   + 2\,{\partial^2 H_0\over\partial j\partial k}\,n_0m_0
   + {\partial^2 H_0\over\partial j^2}\,m_0^2
   = {8\,\mu^2\over\sqrt{1-\mu^2j_0}}
   + {\mu^4k_0\over(1-\mu^2j_0)^{3/2}}\;. \eqno(Omega)
$$
So, to the lowest order in $K$ and $\eps$, the $K$ dynamics near the resonant
torus is generated by
$$
   H_{\rm loc}(\gamma,K) = H_0(k_0,j_0)
   + \half\,\Omega\,K^2 + \ts{\sqrt5\over16}\,\eps^2k_0^2\,j_0
   \cos(\gamma+\psi-\psi_0)\;, \eqno(LocalH)
$$
which is the Hamiltonian of a well-known system, the non-linear pendulum.

Let us now find the homoclinic orbits. If we call $h_0 = H_0(k_0,j_0)$, as
before, and fix some value $H_{\rm loc} = h_{\rm loc}$ for the local
Hamiltonian, we can compute from Eq.\ \(LocalH)
$$
   K = \pm \left\{ {2\over\Omega} \Bigl[ h_{\rm loc}-h_0 - \ts{\sqrt5\over16}
   \,\eps^2k_0^2\,j_0 \cos(\gamma+\psi-\psi_0) \Bigr] \right\}^{1/2}.
$$
To simplify calculations, we set $\psi_0 = \psi+\pi$.
Then, for the  homoclinic orbit, at the maximum $\gamma = \pi$ of the
potential we must have $K = 0$, from which we find that $h_{\rm loc}-h_0
= \ts{\sqrt5\over16}\,\eps^2k_0^2\,j_0$, and
$$
   K = \pm \left\{ \ts{\sqrt5\over8}\,\Omega^{-1}\eps^2k_0^2\,j_0
   \,(1+\cos\gamma) \right\}^{1/2}
   = \pm \left(\ts{\sqrt5\over4}\,\Omega^{-1}\eps^2k_0^2\,j_0\right)^{1/2}
   \cos{\gamma\over2}\;. \eqno(HomoDeltaK)
$$
Finally, we can compute the evolution of $\gamma$ by integrating the Hamilton
equation $\dot\gamma = \partial H_{\rm loc}/\partial K = \Omega\,K$, and we
obtain
$$
   \gamma(t) = 4\arctan\exp\left\{ \left(\ts{\sqrt5\over16}\eps^2\Omega
   \,k_0^2\,j_0 \right)^{\!1/2} t\right\} -\pi\;. \eqno(HomoGamma)
$$
We will use all of these results in the next section.

\section{IV.}{The Melnikov method}\taghead{4.}

\noindent We are now going to study the effect of the other terms of the
perturbation
$$ \eqalignno{
   \delta H^{\prime(2)}(\gamma,\delta;K,J)
   & = \eps^2 \sum_{nm} V_{nm}^{\prime(2)}(k,j)
   \cos(n\theta + m\delta + \psi_{nm}) \cr
   & = \eps^2 \sum_{nm} V_{nm}^{\prime(2)}(k,j) \cos\left({n\over4}\,\gamma
   + {2m-n\over2}\,\delta - {n\over4}\,\psi_0 + \psi_{nm}\right),
   &(otherDeltaH)}
$$
in particular of the terms with $(n,m)\ne(n_0,m_0)$, on the dynamics of the
integable system with Hamiltonian $H_{\rm loc}$, near the homoclinic orbit.
We will use the Melnikov method to show that a stochastic layer forms in the
vicinity of this destroyed separatrix, which acts as a seed for chaos.

We thus want to show\Refto{Guck} that the Melnikov function fo the local $K$
dynamics has a contribution
$$
   M_{nm}(\psi_{nm}):=
   \int_{\rm h.o.}\dd t\,\{H_{\rm loc},\delta H_{nm}^{\prime(2)}\} \eqno(Mel)
$$
which is an oscillating function of $\psi_{nm}$ and has therefore isolated,
transverse zeroes, for some values of $(n,m)$; here, the integral in \(Mel)
is taken over the homoclinic orbit, given by $\gamma(t)$ in \(HomoGamma),
$\delta(t) = \omega_jt + O(\eps^2)$ from $\dot\delta = \de H/\de j'$, $k =
k_0+4K = k_0 + O(\eps)$, and $j = j_0 + O(\eps)$.

In the calculation of the Melnikov integral we will keep only the leading
order terms in $\eps$ and $K$. Then, approximating the Poisson bracket
$$ \eqalignno{
   \{H_{\rm loc},\delta H_{nm}^{\prime(2)}\} &\approx \Bigl\{\half\,
   \Omega K^2,\eps^2V_{nm}^{\prime(2)} \cos\Bigl({n\over4}\,\gamma +
   {2m-n\over2}\,\delta - {n\over4}\,\psi_0 + \psi_{nm}\Bigr) \Bigr\} \cr
   &= \fourth\,\eps^2\Omega K n\,V_{nm}^{\prime(2)}(k_0,j_0) \sin\Bigl(
   {n\over4}\,\gamma + {2m-n\over2}\,\delta - {n\over4}\,\psi_0 +
   \psi_{nm}\Bigr), &(approxPb)}
$$
and using $\dot\gamma = \Omega K$, we get
$$ \eqalignno{
   M_{nm}(\psi_{nm}) &= \int_{-\pi}^\pi\dd\gamma\,\dot\gamma^{-1}\,
   \{H_{\rm loc},\delta H_{nm}^{\prime(2)}\} \cr
   &= \fourth\,\eps^2n\,V_{nm}^{\prime(2)}(k_0,j_0)
   \bigl(A_{nm}\sin\psi'_{nm} + B_{nm}\cos\psi'_{nm}\bigr),&(MelEval)}
$$
where we have defined $\psi'_{nm}:= \psi_{nm}-{n\over4}\psi_0$, and
$$ \eqalignno{
   A_{nm}:&= \int_{-\pi}^\pi\dd\gamma\,
   \cos\Bigl({n\over4}\,\gamma + {2m-n\over2}\,\delta(\gamma)\Bigr)\cr
   B_{nm}:&= \int_{-\pi}^\pi\dd\gamma\,
   \sin\Bigl({n\over4}\,\gamma + {2m-n\over2}\,\delta(\gamma)\Bigr),
   &(AandB)}
$$
and $\delta(\gamma)$ is obtained from \(HomoGamma) and $\delta \approx
\omega_jt$,
$$
   \delta(\gamma) = \eta\,\ln\,\tan{\gamma+\pi\over4}\;,
   \qquad \eta:= \omega_j \left(\ts{\sqrt5\over16}
   \eps^2\Omega\,k_0^2\,j_0 \right)^{\!-1/2}. \eqno(HomoDelta)
$$
The function $M_{nm}$ in \(MelEval) is clearly oscillating whenever
$nV_{nm}^{\prime(2)}A_{nm}$ and/or $nV_{nm}^{\prime(2)}B_{nm}$ are not zero.
We thus have to show that, for some $(n,m)$ in the set $\{(2,0)$, $(2,\pm2)$,
$(2,\pm4)$, $(4,0)$, $(4,-2)$, $(4,\pm4)\}$, $A_{nm}$ or $B_{nm}$ don't
vanish.

Perhaps surprisingly, the integrals in \(AandB) can be calculated
analytically, at least for half of the cases of interest, the ones with $n =
2$. To do this, notice that $A_{nm}$ and $B_{nm}$ are, respectively, the real
and imaginary parts of the complex function
$$ \eqalignno{
   Z_{nm}:&= \int_{-\pi}^\pi\dd\gamma\,\ee^{\ii n\gamma/4}
   \left(\tan{\gamma+\pi\over4}\right)^{\!\ii(2m-n)\eta/2} \cr
   &= 4\,\ee^{-\ii n\pi/4}\int_0^{\pi/2}\dd x\,\ee^{\ii nx}
   (\tan x)^{\ii(2m-n)\eta/2}, &(Znm)}
$$
where we have defined $x:= \fourth\,(\gamma+\pi)$. Therefore, for $n = 2$,
we have\Refto{Grad}
$$ \eqalignno{
   Z_{2m} &= -4\ii \biggl[ \int_0^{\pi/2}\dd x\,\cos2x\,\tan^{\ii(m-1)\eta}x
   + \ii\int_0^{\pi/2}\dd x\,\sin2x\,\tan^{\ii(m-1)\eta}x \biggr] \cr
   &= -4\ii \left[-{\ii(m-1)\eta\pi\over2}\,\sec{\ii(m-1)\eta\pi\over2}
   + {\ii(m-1)\eta\pi\over2}\;{\rm cosec}\,{\ii(m-1)\eta\pi\over2} \right]\cr
   &= -2(m-1)\eta\pi \left[{\rm sech}\,{(m-1)\eta\pi\over2}
   + \ii\;{\rm cosech}\,{(m-1)\eta\pi\over2}\right]; &(Z2m)}
$$
in other words,
$$
   A_{2m} = -2(m-1)\eta\pi\;{\rm sech}\,{(m-1)\eta\pi\over2}\;,\quad
   B_{2m} = -2(m-1)\eta\pi\;{\rm cosech}\,{(m-1)\eta\pi\over2}\;, \eqno(nis2)
$$
which don't vanish for any of the values of $m$ listed above.


\section{V.}{Conclusions}\taghead{5.}

\noindent We have shown analytically that a RW universe, filled with a massive
conformally coupled scalar field, has a chaotic behaviour, at least for a
sufficiently small value of the mass, confirming the numerical results of
\ref{Cal1}. This is just the first step in our program to analyze the
dynamics of RW models with various types of matter content, in order to
relate this to numerical results, and make more precise statements about
physical consequences. Among the purely theoretical aspects of this work, on
the one hand, some possible generalizations to cosmological models differing
by the addition of a cosmological constant and/or other parameters are being
currently studied,\Refto{Bom2} as well as the possibility of adding a
second matter field; on the other hand, we are working on extending the
results obtained for the model in this paper to an estimation of the size of
the stochastic region in phase space and of the time scale for the
manifestation of the chaos we predict. The latter point is an important one
since, as we mentioned in \S2, this time scale must be smaller than the
lifetime of the universe in our model. Although the simulations in \ref{Cal1}
covered many cycles for the universe, it was stated there that chaos would
show up on much smaller time scales, and therefore has, in principle,
observable consequences; in \ref{Corn}, however, the authors argued
qualitatively that they expect the time scale to be much larger than the
lifetime of the universe, and used this to motivate the use of two different
scalar fields in their model. We believe that the issue should be resolved by
an actual, quantitative estimate of the time scale.

We will conclude with some comments on possible avenues for future research,
concerning other aspects of the relationship between chaos and the physics of
the early universe. At a classical level already, chaos opens up ways of
explaining the origin of the arrow of time,\Refto{Cal5} and in our
cosmological setting therefore, of improving our knowledge of cosmological
statistical mechanics and thermodynamics. Furthermore, just as
non-integrability and dynamical instability are, at the classical level, the
causes of chaos, they are also, at the quantum level, the causes of the
instability of quantum states and particles; one therefore expects, for
example, classical chaos and semiclassical particle production to be
related.\Refto{Zure} At the semiclassical level,\Refto{Cast} one can prove
the presence of
decoherence and correlations, which cause the transition to the classical
regime; stable and unstable scalar field
quanta are created during this process. The study of this passage, that leads
from particle creation to chaos is, therefore, imperative. In a model very
similar to ours,\Refto{CCS} a Lyapunov variable was used to define a growing
entropy in the universe; this could be adapted, with minor changes, to the
present model, and looks primising since, for example, growth of entropy,
particle creation, decoherence and isotropization (a tendency to some kind of
equilibrium) are related in semiclassical cosmological models.\Refto{Cal2}

The relevance of chaos in the transition to the semiclassical limit of
quantum cosmology, for RW models, was studied in \ref{Cal4}. At the
complete quantum level, let us recall only that RW models have been used
to study the relation between the cosmological and thermodynamical arrows of
time;\Refto{Hawk} but many other results
related to these subjects can be found in the literature. Ultimately, a very
interesting goal for future work, and possibly a formidable task even in our
simple RW model with chaotic behaviour, is to relate all these fundamental
features of the universe in a unified theoretical framework.


\section{Acknowledgements}{}

\noindent Part of this work was done when the authors were in the RGGR group
at the Universit\'e Libre de Bruxelles in Brussels, Belgium, where it was
partially supported by the Commission of the European Communities, under
contract no.\ ECRU002 (DG-III); we are grateful to Edgard Gunzig for making
this possible, and for the working atmosphere in the group. F.L. would also
like to thank the University of Buenos Aires and the Fundaci\'on Antorchas
for financial support.


\vfill\eject
\section{References}{}\frenchspacing
\leftskip=30pt\parindent=-10pt
\def\new{\hfill\break\indent}

\def\CQG{{\it Class. Quantum Grav. }}
\def\GRG{{\it Gen. Rel. Grav. }}
\def\JMP{{\it J. Math. Phys. }}

\def\PL{{\it Phys. Lett. }}
\def\PRD{{\it Phys. Rev.} D }
\def\PRL{{\it Phys. Rev. Lett. }}

\refis{Berr} M. Berry ``Semiclassical mechanics of regular and irregular
motion'' in {\it Chaotic Behaviour of Deterministic Systems}, edited by G.
Iooss, R.H.G. Helleman and R. Stora (North-Holland, New York, 1983).

\refis{BKL} V.A. Belinskii, Z.M. Khalatnikov and E.M. Lifschitz
``Oscillatory approach to a singular point in the relativistic cosmology''
{\it Adv. Phys.} {\bf19}, 525-73 (1970).

\refis{Bom1} L. Bombelli and E. Calzetta, ``Chaos around a black hole,''
\CQG {\bf9}, 2573-99 (1992);
\new L. Bombelli, ``Particle motion around perturbed black holes: The onset
of chaos'' in {\it Deterministic Chaos in General Relativity}, edited by D.
Hobill, A. Burd and A. Coley (Plenum, New York, 1994).

\refis{Bom2} L. Bombelli and K. Rosquist, ``Families of chaotic cosmologies,''
{\it in preparation} (1998).

\refis{Cal1} E. Calzetta and C. El Hasi, ``Chaotic Friedmann-Robertson-Walker
cosmology,'' \CQG {\bf10}, 1825-41 (1993), and gr-qc/9211027;
\new E. Calzetta, ``Homoclinic chaos in relativistic cosmology,'' in {\it
Deterministic Chaos in General Relativity}, edited by D. Hobill, A. Burd and
A. Coley (Plenum, New York, 1994).

\refis{Cal2} E. Calzetta, ``Particle creation, inflation, and cosmic
isotropy,'' \PRD {\bf44}, 3043-51 (1991);
\new E. Calzetta and F.D. Mazzitelli, ``Decoherence and particle creation,''
\PRD {\bf42}, 4066-9 (1990).

\refis{Cal3} E. Calzetta and C. El Hasi ``Nontrivial dynamics in the early
stages of inflation'' \PRD {\bf51}, 2713-28 (1995), and gr-qc/9408010.

\refis{Cal4} E. Calzetta and J.J. Gonz\'alez ``Chaos and semiclassical
limit in quantum cosmology'' \PRD {\bf51}, 6821-8 (1995), and gr-qc/9411045.

\refis{Cal5} E. Calzetta ``A necessary and sufficient condition for
convergence to equilibrium in Kolmogorov systems'' \JMP {\bf32}, 2903 (1991).

\refis{Cast} M. Castagnino, E. Gunzig and F. Lombardo ``Time asymmetry in
semiclassical cosmology'' \GRG {\bf27}, 257-66 (1995);
\new M. Castagnino and F. Lombardo ``Decoherence, correlations, and unstable
quantum states in semiclassical cosmology'' \GRG {\bf28}, 263 (1996), and
gr-qc/9404028.
\new F. Lombardo and F. Mazzitelli ``Einstein-Langevin equations from running
coupling constants'' \PRD {\bf55}, 3889-92 (1997), and gr-qc/9609073.

\refis{CCS} E. Calzetta, M. Castagnino and R. Scoccimarro, \PRD {\bf45},
2806 (1992).

\refis{Corn} N.J. Cornish and J.J. Levin, ``Chaos, fractals and inflation''
\PRD {\bf53}, 3022-32 (1996), and astro-ph/9510010.

\refis{Fine} F. Finelli, G.P. Vacca and G. Venturi, ``Chaotic inflation with a
scalar field in non-classical states,'' gr-qc/9712098

\refis{Gall} G. Gallavotti, {\it The Elements of Mechanics\/} (Springer, New
York, 1983).

\refis{Grad} I.S. Gradshteyn and I.M. Ryzhik, {\it Tables of Integrals,
Series, and Products\/} (Academic Press, 1980).

\refis{Guck} J. Guckenheimer and P. Holmes, {\it Nonlinear Oscillations,
Dynamical Systems, and Bifurcations of Vector Fields\/} (Springer, Berlin,
1983).

\refis{Gutz} M.C. Gutzwiller, {\it Chaos in Classical and Quantum
Mechanics\/} (Springer, New York, 1990).

\refis{Hawk} S.W. Hawking, ``Arrow of time in cosmology,'' \PRD {\bf32},
2489-95 (1985);
\new D.N. Page, ``Will entropy decrease if the universe recollapses?'' \PRD
{\bf32}, 2496-9 (1985).

\refis{Misn} C.W. Misner, ``Mixmaster universe,'' \PRL {\bf22}, 1071-4 (1969).

\refis{MTW} C.W. Misner, K.S. Thorne and J.A. Wheeler, {\it Gravitation}
(Freeman, 1973).

\refis{Ozor} A.M. Ozorio de Almeida, {\it Hamiltonian Systems: Chaos
and Quantization\/} (Cambridge University Press, Cambridge, 1988).

\refis{Pull} J. Pullin, ``Time and chaos in general relativity,'' in {\it
Relativity and Gravitation: Classical and Quantum}, edited by J.C. D'Olivo et
al.\ (World Scientific, Singapore, 1991);
\new A. Burd and R. Tavakol, ``Invariant Lyapunov exponents and chaos in
cosmology,'' \PRD {\bf47}, 5336-41 (1993)
\new S.E. Rugh, ``Chaos in the Einstein equations -- Characterization and
importance,'' in {\it Deterministic Chaos in General Relativity}, edited by
D. Hobill, A. Burd and A. Coley (Plenum, New York, 1994);
\new B.K. Berger, D. Garfinkle and E. Strasser, ``New algorithm for
mixmaster dynamics,'' \CQG {\bf14}, L29-36 (1997), and gr-qc/9609072.

\refis{Vuce} A. Helmi and H. Vucetich, ``Non-integrability and chaos in
classical cosmology,'' \PL A {\bf230}, 153-6 (1997), and gr-qc/9705009.

\refis{Zasl} G.M. Zaslavsky, R.Z. Sagdeev, D.A. Usikov and A.A. Chernikov,
{\it Weak Chaos and Quasi-Regular Patterns\/} (Cambridge University Press,
Cambridge, 1991).

\refis{Zure} W.H. Zurek and J.P. Paz, ``Decoherence, chaos, and the
second law,'' \PRL {\bf72}, 2508-11 (1994), and gr-qc/9402006.

\endreferences
\endit
\end